\documentclass[aps,preprint,nofootinbib,floatfix]{revtex4}
\usepackage{graphicx,color}
\usepackage[latin1]{inputenc}
\usepackage{amsmath,amssymb}
\usepackage{hyperref}
\usepackage{epstopdf}
\usepackage{slashed}
\usepackage{soul}
\usepackage{amssymb}
\usepackage[normalem]{ulem}
\usepackage[caption=false]{subfig}
\usepackage{booktabs}
\usepackage{diagbox}
\usepackage{makecell}
\usepackage{multirow}
\usepackage{siunitx}
\newcommand{\lsim}{\mathrel{\mathop{\kern 0pt \rlap
  {\raise.2ex\hbox{$<$}}}
  \lower.9ex\hbox{\kern-.190em $\sim$}}}
\newcommand{\gsim}{\mathrel{\mathop{\kern 0pt \rlap
  {\raise.2ex\hbox{$>$}}}
  \lower.9ex\hbox{\kern-.190em $\sim$}}}

\newcommand{\sv}{\ensuremath{\langle\sigma v\rangle}}

\newcommand{\mev}{\ensuremath{\,\mathrm{MeV}}}
\newcommand{\gev}{\ensuremath{\,\mathrm{GeV}}}

\def  \bcen   {\begin{center}}
\def  \ecen   {\end{center}}
\def  \beq    {\begin{equation}}
\def  \eeq    {\end{equation}}
\def  \beqa   {\begin{eqnarray}}
\def  \eeqa   {\end{eqnarray}}

\def\bea{\begin{eqnarray}}
\def\eea{\end{eqnarray}}

\begin{document}

\title{Prospects for Probing Sub-GeV Leptophilic Dark Matter with the Future VLAST}
\author{Tian-Peng Tang$^{a}$}
\author{Meiwen Yang$^{c,a}$}
\author{Kai-Kai Duan$^{a}$}
\author{Yue-Lin Sming Tsai$^{a,b}$}
\author{Yi-Zhong Fan$^{a,b}$}

\affiliation{$^a$Key Laboratory of Dark Matter and Space Astronomy, 
   Purple Mountain Observatory, Chinese Academy of Sciences, Nanjing 210033, China}
\affiliation{$^b$School of Astronomy and Space Science, University of Science and Technology of China, Hefei, Anhui 230026, China}
\affiliation{$^c$Department of Physics and Institute of Theoretical Physics, 
Nanjing Normal University, Nanjing, 210023, China }

\date{\today}

\begin{abstract} 
The proposed Very Large Area Space Telescope (VLAST), with its expected unprecedented sensitivity in the MeV-GeV range, can also address the longstanding ``MeV Gap" in gamma-ray observations. 
We explore the capability of VLAST to detect sub-GeV leptophilic dark matter (DM) annihilation, focusing on scalar and vector mediators and emphasizing the resonance region where the mediator mass is approximately twice the DM mass. 
While $s$-wave annihilation is tightly constrained by relic density and cosmic microwave background observations, 
$p$-wave and mixed $(s+p)$-wave scenarios remain viable, particularly near resonance. 
Additionally, direct detection experiments, especially those probing DM-electron scattering, significantly constrain nonresonance parameter space but are less effective in the resonance regime. 
VLAST can uniquely probe this surviving region, outperforming existing and planned instruments, and establishing itself as a crucial tool for indirect detection of thermal relic DM. 
\end{abstract}

\maketitle

\section{Introduction \label{sec:intro}}

The search for non-gravitational signals from dark matter (DM) is essential to understanding its fundamental properties, 
particularly its interactions with Standard Model (SM) particles. 
Within the thermal relic paradigm, DM masses between the MeV and TeV scales can naturally reproduce the observed abundance through freeze-out dynamics~\cite{Griest:1989wd,Boehm:2003bt,Boehm:2002yz}. 
This framework has long motivated the canonical WIMP program in the GeV to TeV range, 
where extensive experimental efforts have been carried out~\cite{Chang:2008aa,HESS:2009chc,PAMELA:2008gwm,Hooper:2010mq,Zhou:2014lva,Calore:2014xka,Daylan:2014rsa,Cui:2016ppb,Cuoco:2016eej,Cholis:2019ejx,Fan:2022dck,Fan:2024rcr,Feng:2008ya,Arcadi:2017kky,Cirelli:2024ssz}. 
Although intriguing anomalies have been reported, a wide array of direct, indirect, and collider searches has severely constrained large portions of the viable parameter space above the GeV scale.

In contrast, thermal relic DM with masses below a GeV is comparatively less explored but still theoretically well motivated. 
The very light region, at or below the MeV scale, is strongly constrained by cosmological probes such as Big Bang Nucleosynthesis (BBN) and the Cosmic Microwave Background (CMB), leaving only a narrow mass window. 
The intermediate regime between the MeV and GeV scales, however, remains open~~\cite{Adams:1998nr,Chen:2003gz,Padmanabhan:2005es,Slatyer:2009yq,Chluba:2011hw,Slatyer:2015jla,Steigman:2015hda,Planck:2018nkj,Cang:2020exa,Kawasaki:2021etm,Matsumoto:2022ojl}, 
even though existing limits from beam dump experiments, low-threshold direct detection, and CMB observations already exclude scenarios with large couplings to the SM particles.

This situation makes the sub-GeV mass range particularly appealing. 
It arises naturally within the thermal relic framework, has not yet been excluded by current data, and offers a parameter space that is both finite and testable. 
With upcoming advances across multiple experimental frontiers, the remaining window of sub-GeV DM can be probed decisively in the near future~\cite{Boddy:2015efa,Bartels:2017dpb,Matsumoto:2018acr,Chen:2018vkr,Cirelli:2020bpc,Coogan:2021sjs,Caputo:2022dkz,Guo:2023kqt,Berlin:2023qco,DelaTorreLuque:2023olp,Chen:2024njd,ODonnell:2024aaw,Balan:2024cmq,Nguyen:2024kwy,Saha:2025wgg,Cirelli:2025qxx,Watanabe:2025pvc}. 
It thus provides a well-motivated and experimentally timely target for the next generation of DM indirect searches.

Historically, instruments like COMPTEL~\cite{Strong:1998ck} and EGRET~\cite{Hunter:1997qec} conducted gamma-ray observations 
in the range of $\mathcal{O}(1-20)~\mev$. 
However, their sensitivities were several orders of magnitude lower than those of higher-energy experiments such as Fermi-LAT~\cite{Fermi-LAT:2009ihh} and DAMPE~\cite{DAMPE:2017cev}, leaving an observational gap known as the ``MeV Gap". 
In recent years, several next-generation gamma-ray telescopes have been proposed to close this gap and enhance sensitivity to MeV-scale gamma rays, 
including COSI~\cite{Tomsick:2021wed}, the Very Large Area gamma-ray Space Telescope (VLAST)~\cite{2022AcASn..63...27F,Pan:2024adp}, 
e-ASTROGAM~\cite{e-ASTROGAM:2017pxr},  GECCO~\cite{Orlando:2021get}, AMEGO~\cite{Kierans:2020otl}, GRAMS~\cite{Aramaki:2019bpi}, 
AdEPT~\cite{Hunter:2013wla}, and MAST~\cite{Dzhatdoev:2019kay}.

VLAST~\cite{2022AcASn..63...27F,Pan:2024adp} is a space mission designed to detect gamma-ray photons through both Compton scattering and electron-positron pair production mechanisms.
The mission aims to perform a comprehensive all-sky survey from low-Earth orbit using three primary detector components: an anti-coincidence detector to suppress charged-particle background, a Silicon tracker coupled with a low-energy gamma-ray detector for accurate reconstruction of particle trajectories and energy measurement at low energies, and a high-energy imaging calorimeter to measure energy deposition and reconstruct the electromagnetic showers induced by high-energy particles.
VLAST builds upon the foundational principles and structure of previous gamma-ray detector but introduces a key modification which replaces the tungsten converter in the tracker with thin CsI tiles. This modification enables energy measurement of Compton electrons, along with simultaneous energy and direction measurements of Compton photons, thereby allowing precise reconstruction of Compton events, especially for the photon energy around $\mathcal{O}(\mev)$.
This detector configuration enables simultaneous observation across a broad energy range from MeV to TeV, supporting key scientific objectives such as searching for DM signatures, monitoring of transient and persistent astrophysical phenomena, probing cosmic ray origins and transportation, studying cosmic evolution, and testing fundamental physical laws.

Given the capabilities of VLAST in the MeV to GeV range, leptonic annihilation channels provide especially well-motivated benchmarks. 
They offer three advantages. 
First, leptonic final states produce cleaner and more robust gamma-ray spectra than hadronic channels, 
which suffer from QCD uncertainties and model dependence in the 100~MeV to GeV DM mass range~\cite{Plehn:2019jeo}. 
Second, their interactions are subject to complementary laboratory probes: electron couplings are constrained by low-threshold direct detection experiments such as XENON1T~\cite{XENON:2019gfn}, DarkSide-50~\cite{DarkSide:2022knj}, DAMIC-M~\cite{DAMIC-M:2023gxo}, and SENSEI~\cite{SENSEI:2020dpa}, while muon couplings are increasingly tested by collider and beam-dump searches including NA64$\mu$~\cite{2409.10128}, Belle II~\cite{2403.02841}, and BABAR~\cite{1606.03501}. 
Third, they are motivated by well-defined theoretical frameworks, including anomaly-free $L_i-L_j$ gauge symmetries~\cite{Foot:1990mn,Foot:1994vd}, scalar or pseudoscalar mediators with dominant leptonic couplings~\cite{Fox:2008kb}, 
and leptophilic dark photons or axion-like particles~\cite{Fayet:1990wx,Fayet:1980rr,Calibbi:2016hwq,Han:2020dwo}. 
Altogether, these considerations make leptophilic DM a theoretically clean and experimentally timely target for future gamma-ray telescopes such as VLAST.

In this work, we investigate the sensitivity of VLAST to sub-GeV leptophilic DM annihilation in DM-rich dwarf galaxy, such as Draco.
Based on the systematic construction and classification of 16 renormalizable effective leptophilic models~\cite{Abdughani:2021oit},
we focus on three representative DM interactions:
(i) scalar DM with a scalar mediator, (ii) Dirac DM with a scalar mediator, and (iii) Dirac DM with a vector mediator.
These interactions exhibit distinct annihilation behaviors: $s$-wave, $p$-wave, and mixed ($s+p$)-wave processes, respectively.
A comprehensive Markov Chain Monte Carlo (MCMC) analysis is performed, incorporating constraints from DM relic abundance, CMB observations, BBN, DM self-interactions, direct detection experiments, and beam dump experiments. 
Our analysis shows that direct detection experiments like XENON1T impose stringent limits on DM-electron interactions, particularly in the electrophilic scenario. 
VLAST, however, holds strong potential to probe large regions of the $p$-wave and $(s+p)$-wave annihilation parameter space. 
Surprisingly, the resonance region offers a promising goal for VLAST. 
In contrast, the $s$-wave scenario remains challenging due to stringent constraints from CMB observations and relic density requirements.

The remainder of this paper is organized as follows.
In Sec.~\ref{sec:Vlast}, we briefly present the key performance characteristics of VLAST and estimate its sensitivity to sub-GeV DM.
In Sec.~\ref{sec:models and constraints}, we provide an overview of the leptophilic DM Lagrangians considered in this work and outline the constraints incorporated into our likelihood function analysis.
The allowed parameter space resulting from the likelihood analysis is presented in Sec.~\ref{sec:result}.
Finally, we summarize our conclusions in Sec.~\ref{sec:summary}.

\section{The VLAST sensitivity prospects}
\label{sec:Vlast}

The VLAST detector configuration has been proposed~\cite{2022AcASn..63...27F,Pan:2024adp}, and its performance has been evaluated through Monte Carlo simulations using the \texttt{GEANT4} toolkit, confirming the feasibility of the instrument. 
VLAST provides an acceptance greater than $10~\mathrm{m}^2~\mathrm{sr}$ (four times that of Fermi-LAT) and achieves energy and angular resolutions of better than 2\% and 0.2$^{\circ}$ at 10~GeV, respectively. 
These features significantly enhance its sensitivity to sub-GeV leptophilic DM annihilation in DM-rich systems, such as Draco dwarf galaxy.


Draco dwarf galaxy has an angular size of $\sim 0.26^\circ$~\cite{2016JCAP...02..039M}, 
much smaller than the angular resolution of VLAST in the MeV range ($\sim 5^\circ$ at 1 MeV and $\sim 3^\circ$ at 10 MeV). 
Therefore, any DM-induced $\gamma$-ray emission from Draco can be treated as a point source for VLAST. 
We calculate the instrument sensitivity to a point source at Draco location and translate it into limits on the DM annihilation cross section. 
Previous $\gamma$-ray dwarf galaxy studies~\cite{Fermi-LAT:2015att, Fermi-LAT:2016uux} have shown that sensitivity can be computed for individual energy bins and is independent of the spectral index. 
Following their approach, to simulate the expected signal, 
a point source with a power-law spectrum of the form $d\Phi_{\rm PS}/dE_\gamma \propto E_\gamma^{-2}$ is used as the baseline model for sensitivity calculations.

The astrophysical background model contains the Galactic Diffuse Emission (GDE) and the Cosmic $\gamma$-ray Background (CGB).  
The GDE component arises from inverse Compton scattering, bremsstrahlung, and $\pi^0$ decay, and 
is modeled with \texttt{GALPROP} following the method of Ref.~\cite{Fermi-LAT:2012edv}.  
Similarly, following COMPTEL~\cite{2000AIPC..510..467W}, the CGB component is modeled as a broken power law
\begin{equation}
    \frac{d\Phi_{\rm CGB}}{dE_\gamma}\left(E_\gamma\right) = 
    2.2 \times 10^{-4} \left(\frac{E_\gamma}{3~\text{MeV}} \right)^{-\Gamma}~(\text{MeV cm}^2~\text{s sr})^{-1},
\end{equation}
with photon indices $\Gamma=3.3$ for $E_\gamma\leq 3~\mathrm{MeV}$ and $\Gamma=2.0$ for $E_\gamma > 3~\mathrm{MeV}$.
At GeV energies, we adopt the \texttt{Fermi-LAT Model A} spectrum~\cite{Fermi-LAT:2014ryh}. 
All-sky maps for each component are obtained from publicly available templates
\footnote{\url{https://tsuji703.github.io/MeV-All-Sky/files/allsky/Table_allsky.html}}.


The simulated sky maps include the GDE, CGB, and the point source. 
The counts are generated using VLAST energy-dependent effective area and point-spread function. 
A binned likelihood analysis is performed, 
where the normalizations of GDE and CGB are treated as nuisance parameters. 

The test statistic (TS) is defined as 
\begin{equation}
    {\rm TS} = -2 \ln \left( \frac{L_{\text{GDE+CGB}}}{L_{\text{GDE+CGB+PS}}} \right), 
\end{equation}
where $L_{\text{GDE+CGB+PS}}$ and $L_{\text{GDE+CGB}}$ are the likelihoods with and without the point source, respectively. 
By varying the point source normalization until the TS reaches 25 (corresponding to $5\sigma$ significance), 
we determine the point-source sensitivity of VLAST for one year of survey observations.

\begin{figure}[ht!]
	\centering
    \includegraphics[width=7.5cm,height=7.5cm]{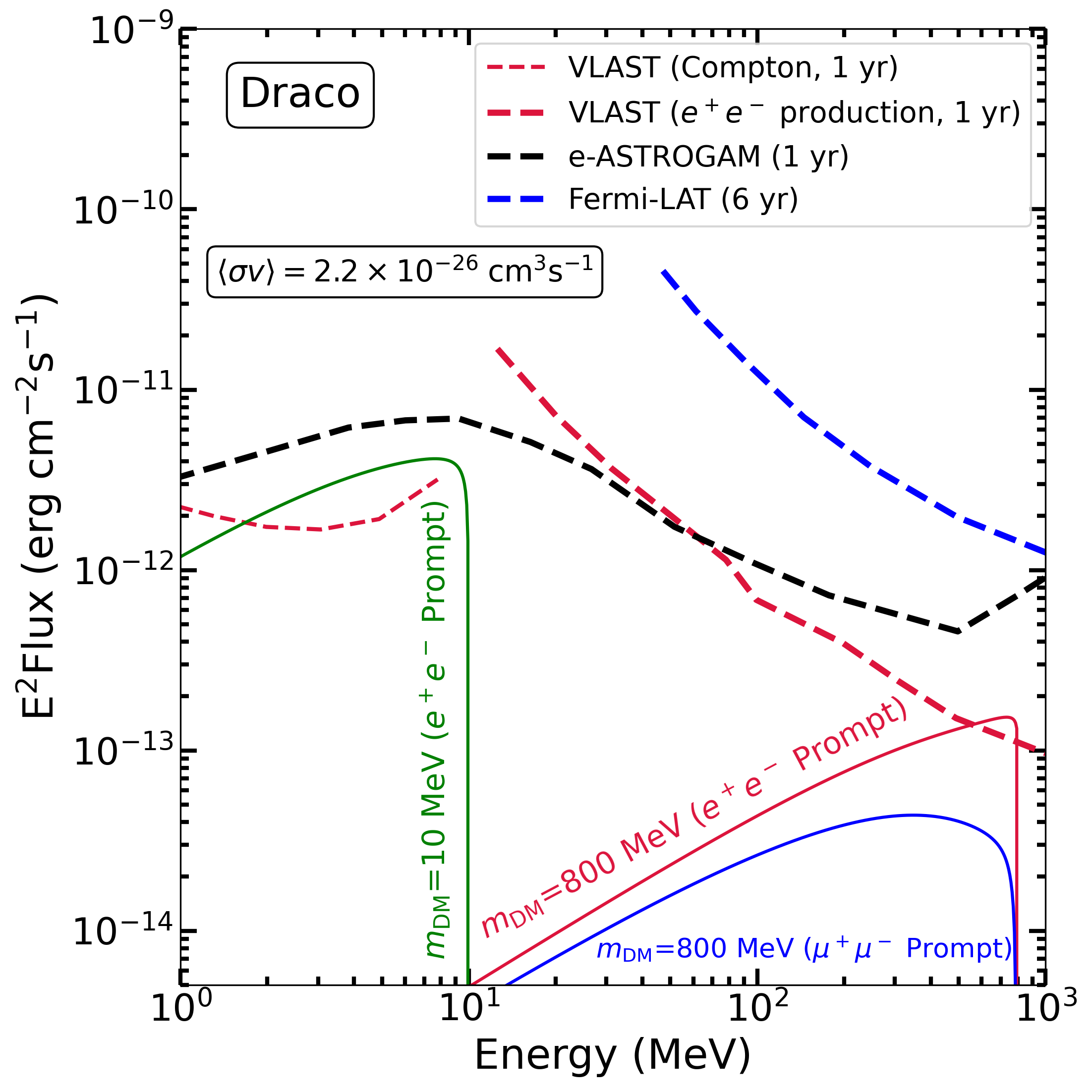}
	\includegraphics[width=7.5cm,height=7.5cm]{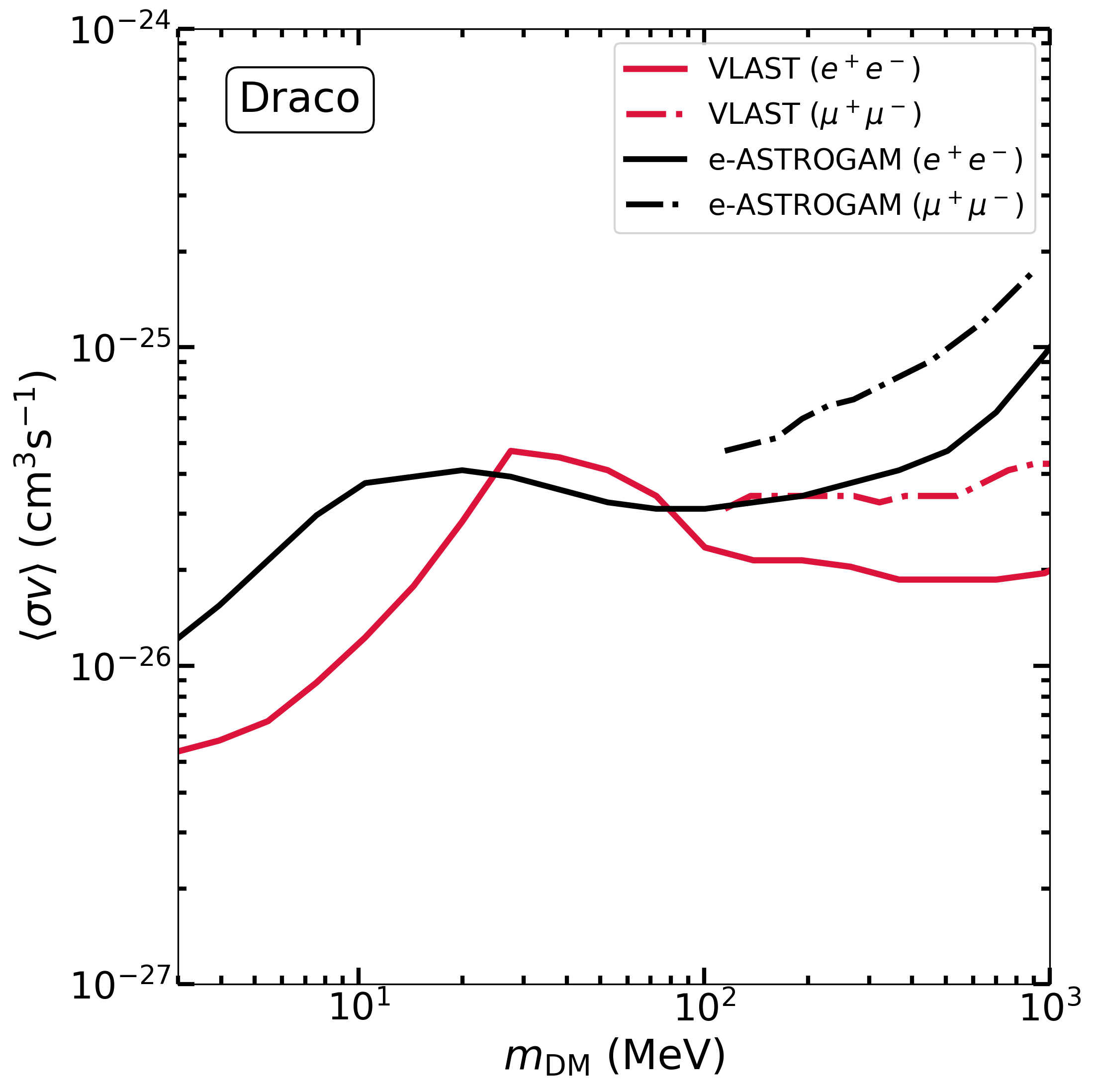}
\caption{
\textbf{Left panel:} The sensitivity of VLAST for one year (red dashed line) to gamma-ray emission from Draco dwarf galaxy, 
compared with projected sensitivities of e-ASTROGAM for one year (black dashed line)~\cite{e-ASTROGAM:2017pxr} and 
6-year Fermi-LAT data (blue dashed line)~\cite{Fermi-LAT:2015att, Fermi-LAT:2016uux}. 
Solid lines show the prompt gamma-ray spectra from DM annihilation into lepton pairs assuming a thermal relic cross section of $\langle \sigma v \rangle = 2.2 \times 10^{-26}~\rm cm^3\,s^{-1}$~\cite{Steigman:2012nb} with $\mathrm{DM}+\mathrm{DM} \to e^+e^-$ for two benchmark masses $m_{\rm DM} = 10~\mev$ (green) and $800~\mev$ (red), and $\mathrm{DM}+\mathrm{DM} \to \mu^+\mu^-$ for $m_{\rm DM} = 800~\mev$ (blue).
\textbf{Right panel:} Projected sensitivities to the DM annihilation cross section. 
Red lines correspond to VLAST sensitivities for $e^+e^-$ (solid) and $\mu^+\mu^-$ (dash-dotted) channels, 
while black lines show the corresponding sensitivities for e-ASTROGAM.
}
	\label{Fig:vlast_sensitivity}
\end{figure}

The left panel of Fig.~\ref{Fig:vlast_sensitivity} shows the projected one-year sensitivity of VLAST (red dashed line) and e-ASTROGAM~\cite{e-ASTROGAM:2017pxr} (black dashed line), compared to 6-year Fermi-LAT data~\cite{Fermi-LAT:2015att, Fermi-LAT:2016uux} (blue dashed line), 
for DM annihilation in Draco dwarf galaxy. 
The right panel presents the corresponding upper limits on the annihilation cross section. 
Note that VLAST shows a sensitivity gap around $10\mev$, resulting from the transition between its two detection mechanisms: Compton scattering and pair production. 
This transition causes discontinuities in the effective area and angular resolution near $10\mev$, leading to the observed dip in sensitivity.

To highlight the VLAST capability to detect MeV gamma-ray signals from DM annihilation,
we compare theoretical flux predictions from Draco dwarf galaxy, known for its clean astrophysical background. 
For the DM density distribution, we adopt Navarro-Frenk-White  (NFW) profile, 
\begin{equation}
    \rho_{\rm NFW}(r) = \frac{\rho_s}{(r/r_s)(1+r/r_s)^2}.
\end{equation}
For Draco, following Ref.~\cite{Pace:2018tin}, we use the scaled density $\rho_s = 2.96\,\gev\mathrm{/cm}^3$, the scaled radius $r_s = 728\,\rm pc$, and the distance $D = 76\,\rm kpc$ from Earth to its center.
Therefore, the differential DM-induced gamma-ray flux is given by 
\begin{equation}\label{VD_Eq2}
\begin{aligned}
\frac{d\phi_{\gamma}}{dE_{\gamma}} &=
\int_{0}^{D} \frac{\rho^2_{\rm NFW}(r) r^2}{4 m_\chi^2 D^2}
\frac{dN_\gamma}{dE_\gamma}  \langle \sigma v_{\rm rel.} (r) \rangle  dr ,\\
\end{aligned}
\end{equation}
 where $dN_\gamma/dE_\gamma$ is the photon yield spectrum, obtained using Hazma~\cite{Coogan:2019qpu}.
 The velocity averaged annihilation cross section $\langle \sigma v_{\rm rel.}(r)\rangle$ depends on the relative velocity distribution of DM particles at radius $r$, given by
\begin{equation}
\begin{aligned}
    &\langle \sigma v_{\rm rel} (r)\rangle=\frac{1}{Norm(r)}\int_0^{2v_{\rm esc}}
\sigma v_{\rm rel} f_{v_{\rm rel.}}(r,v_{\rm rel.})  dv_{\rm rel}.
\end{aligned}
\end{equation}
Here, $v_{\rm esc}$ is the escape velocity, and $v_{\rm rel}$ is the relative velocity.  
The radius-dependent factor $Norm(r)$ normalizes the relative velocity distribution function $f_{v_{\rm rel.}}(r, v_{\rm rel.})$ during DM annihilation, 
\begin{equation}
    \begin{aligned}
       & f_{v_{\rm rel.}}(r,v_{\rm rel.})= v_{\rm rel}^2 
\int_0^{v_{\rm c}} v_{\rm CM}^2 dv_{\rm CM} 
 \int_{-\mu_0}^{\mu_0} d\cos\theta \; 
\; f(v_1,r)f(v_2,r),\\
&{\rm where} \quad \mu_0  =\frac{4 v_{\rm esc}^2- 4 v_{\rm CM}^2-v_{\rm rel.}^2}{4 v_{\rm rel.}v_{\rm CM}}, \quad {\rm and} \quad v_{\rm c} = \sqrt{1-E_{\rm cm}^2/E_{\rm esc}^2}.
    \end{aligned}
\end{equation} 
The two DM particles have velocities $v_1$ and $v_2$, with an angle $\theta$ between them. 
The center-of-mass velocity is $v_{\rm CM}$, the center-of-mass energy is $E_{\rm cm}$, 
and $E_{\rm esc}$ denotes the total energy when both particles are at escape velocity.
For $f(v,r)$, we assume a Maxwell-Boltzmann distribution 
\begin{equation}
f(v,r) = \frac{1}
{\left[2\pi v_{\rm d}^2(r)\right]^{3/2}} 
\exp\left(\frac{-v^2}{2 v_{\rm d}^2(r)}\right)
\end{equation}
where $v_{\rm d}(r)$ is the velocity dispersion  derived from Jeans equation \cite{Jin:2003sj}.

In the left panel of Fig.~\ref{Fig:vlast_sensitivity}, we display the prompt emission from DM annihilation into $e^+e^-$ and $\mu^+\mu^-$ channels, 
assuming a representative cross section of $\langle \sigma v \rangle = 2.2 \times 10^{-26}\,\rm cm^3\,s^{-1}$~\cite{Steigman:2012nb} for two benchmark DM masses, $m_{\rm DM} = 10\,\mev$ and $800\,\mev$. The resulting spectra are shown as solid lines. 
Note that the DM-induced gamma-ray flux due to inverse Compton scattering with CMB photons is not depicted, as it lies in the X-ray range, 
below the threshold of VLAST.

The right panel of Fig.~\ref{Fig:vlast_sensitivity} illustrates the projected sensitivities to the annihilation cross section $\langle \sigma v \rangle$ from null observation of Draco. 
The red and black curves indicate the sensitivities of VLAST and e-ASTROGAM, respectively, for the ${\rm DM + DM} \to e^+ + e^-$ channel (solid lines) and the ${\rm DM + DM} \to \mu^+ + \mu^-$ channel (dash-dotted lines). 
Unlike flux limits, the upper bound on $\sv$ shows a continuous feature around $30\mev$. 
This arises because the most stringent flux constraints, whether from gamma ray energies above or below $10\mev$, 
combine to produce a connected limit near this mass scale.
The VLAST sensitivity demonstrates improved limits compared to e-ASTROGAM, with its performance surpassing e-ASTROGAM by a factor of 2 to 5. 
This enhanced sensitivity is particularly presented in the energy ranges of $1-10 \, \mev$ and $100 \, \mev - 1 \, \gev$.

\section{Sub-GeV DM interactions and their constraints}
\label{sec:models and constraints}
In the sub-GeV mass range, DM particles can annihilate into electrons, muons, and light quarks. 
Given that leptophilic portal DM models are commonly studied in this region, 
we focus on DM-electron and DM-muon interactions in this work. 
We first summarize the selected leptophilic effective DM models in Sec.~\ref{sec:models}, 
followed by an overview of the relevant experimental constraints and likelihoods in Sec.~\ref{sec:constraints}.

\subsection{Benchmark models}
\label{sec:models}

In Ref.~\cite{Abdughani:2021oit}, all representative interactions for muonphilic DM models were systematically classified. 
These models include DM and mediator (MED) particles, which can be scalar, fermion, or vector.
To prevent DM decay, we introduced a $Z_2$ symmetry and constructed 16 renormalizable effective models with $Z_2$-even mediators, 
as summarized in Table II of Ref.~\cite{Abdughani:2021oit}.

In this study, we first perform a preliminary scan of these 16 models and categorize them into three groups based on their annihilation characteristics: $s$-wave, $p$-wave, and $s$+$p$-wave.
Moreover, we find that models exhibiting the same annihilation behavior share highly similar surviving parameter spaces after incorporating the experimental constraints listed in Table~\ref{Tab:constraints}.
Given these findings, we select three benchmark models as the focus of our analysis and presentation: scalar DM with a scalar mediator, Dirac DM with a scalar mediator, and  Dirac DM with a vector mediator.
The corresponding Lagrangians for these models are denoted as $\mathcal{L}_{SS}$, $\mathcal{L}_{FS}$, and $\mathcal{L}_{FV}$, respectively.
The detailed discussions of these models is presented in the following section.

For DM particles promptly annihilating into detectable $\gamma$-rays with energy $\mathcal{O}(10\ \text{MeV})$, 
their mass must be in the MeV to GeV range. 
Additionally, studies~\cite{Matsumoto:2018acr,Chen:2024njd, Chen:2018vkr} indicate that the constraints of the thermal relic density requires 
sub-GeV DM annihilation via a mediator with mass below GeV. 
For leptophilic mediators, such a light particle can only decay into muons or electrons. 
Therefore, to focus on leptophilic mediators, we individually study two annihilation channels: 
(i) ${\rm DM}+{\rm DM} \to e^+ + e^-$ or ${\rm DM}+{\rm DM} \to 4e$, and 
(ii) ${\rm DM}+{\rm DM} \to \mu^+ + \mu^-$ or ${\rm DM}+{\rm DM} \to 4\mu$. 
Based on these motivations, we list three representative interactions.
\begin{itemize}
\item[(i)] \textbf{Scalar DM and scalar mediator:}\\
Consider a real scalar mediator $\phi$ with mass below GeV, which couples both to the scalar dark matter (DM) particle $S$ and 
the SM fermions $f$. The Lagrangian is
\begin{equation}
    \mathcal{L}_{SS} = M_D S^\dagger S \phi + g_f \bar{f} f \phi, 
    \label{eq:model1}
\end{equation}
where $f$ can be either $e^\pm$ or $\mu^\pm$ in this work.   
The dimensional coupling between $S$ and $\phi$ is $M_D$, 
while $g_f$ denotes the coupling between the mediator and the fermions.
This interaction suggests that \textit{DM annihilation proceeds via $s$-wave}, 
with a cross section around $\mathcal{O}(10^{-26})$~cm$^{-3}$s$^{-1}$ to produce the correct relic density $\Omega h^2 \approx 0.1$. 
However, such $s$-wave annihilation with DM masses below $1\gev$ may inject excessive energy, 
potentially distorting the temperature and polarization spectra of the CMB~\cite{Matsumoto:2018acr,Chen:2024njd,Wang:2025tdx}. 
Therefore, CMB observations impose stringent constraints on the coupling constants $g_D$ and $g_f$, particularly for sub-GeV DM $s$-wave annihilation.

\item[(ii)] \textbf{Dirac DM and scalar mediator:}\\
The Lagrangian for the interaction where fermionic (spin-$1/2$) DM $\chi$ couples directly to the lepton sector via a scalar mediator $\phi$ is  
 \begin{equation}
 \mathcal{L}_{FS}=g_D\bar{\chi}\chi\phi+g_f\bar{f}f\phi.
 \label{eq:model2}
\end{equation}
This interaction leads to \textit{$p$-wave annihilation}, suppressed at low velocities ($\sv \propto v_{\rm rel.}^2$). 
As a result, it bypasses CMB constraints, which correspond to $v_{\rm rel.} \approx 10^{-8}$ of the speed of light, while still producing the correct relic density. 
In the present universe, this suppression remains sufficient, with DM relative velocities around $10^{-4}$ to $10^{-3}$ of the speed of light. 
However, thanks to the improved sensitivity of gamma-ray telescopes, $p$-wave annihilation may produce detectable signals in DM-rich regions such as dSphs, especially when accompanied by a resonance condition.

 \item[(iii)] \textbf{Dirac DM and vector mediator:}\\
For Dirac DM annihilation mediated by a vector field $V_\mu$, the Lagrangian is given by
\begin{equation}
\mathcal{L}_{FV} = g_D \bar{\chi} \gamma^\mu \gamma^5 \chi V_\mu + g_f \bar{f} \gamma^\mu \gamma^5 f V_\mu.
\label{eq:model3}
\end{equation}
The annihilation process exhibits \textit{both $s$-wave and $p$-wave contributions}, i.e., $(s + p)$-wave. 
At low velocities, $s$-wave annihilation dominates, leading to a constant annihilation rate similar to that in Eq.~\eqref{eq:model1} (scalar DM and mediator). 
However, at higher velocities, the $p$-wave component becomes dominant, enhancing the annihilation rate. 
This mixed behavior can be probed through both CMB observations and high-energy DM indirect detection.

\end{itemize}

\subsection{Experimental Constraints}
\label{sec:constraints}

\begin{table}[htp]
\small
\centering
\begin{tabular}{|c|c|c|c|c|}
\hline\hline
      Observable & Likelihood type & Experimental data  \\
\hline\hline
\makecell[c]{ Relic abundance} 
      & Gaussian  &  \makecell[c]{$ \Omega_{\rm{DM}}^{\rm exp} h^{2} =0.1193\pm 0.0014$~\cite{Planck:2018vyg};\\ 
      $\sigma_{\rm sys}= 10\%\times \Omega_{\rm{DM}}^{\rm theo} h^{2}$.} \\
\hline
\makecell[c]{DM direct detection}
& Half-Gaussian 
&\makecell[c]{
$ 60 \mev < m_{\rm{DM}} < 1\gev$ (DarkSide-50~\cite{DarkSide-50:2023fcw}),\\
$ 160 \mev < m_{\rm{DM}} < 1\gev$ (CRESST~\cite{CRESST:2019jnq,CRESST:2022lqw}).
}  
\\
\hline
\makecell[c]{ CMB }
      & Half-Gaussian
      &{$\sv^{\text{Planck}}_{\text{CMB}}$ for 95\% C.L.}~\cite{Planck:2018vyg} \\
\hline
\makecell[c]{DM Self Interaction}
&Half-Gaussian&$\sigma_{\rm{DM},\rm{DM}\to\rm{DM},\rm{DM}}/m_{\text{DM}}<1.0~{\rm cm^2/g}$~\cite{Randall:2008ppe}\\
\hline
\makecell[c]{$\triangle N_{\rm eff}$}
&Half-Gaussian&$\triangle N_{\rm eff}< 0.17$ for 95\% C.L.~\cite{Planck:2018vyg}\\
\hline
BBN
& Conditions  &\makecell[c]{
$\tau_{\text{MED}}\le 10^{5}$~s~\cite{Krnjaic:2015mbs}
 }  \\
\hline
Mediator production
& Conditions  &\makecell[c]{
See Refs.~\cite{2502.04053, 2409.10128,2403.02841,1606.03501,NA64:2021xzo} for details.
}  \\ 
\hline\hline

\end{tabular}
\caption{\label{Tab:constraints}
Summary of observables, likelihood function types, and experimental data used in this work.
}
\end{table}

In this section, we summarize the experimental and observational data, 
and explain how they are incorporated into the likelihood functions used in our MCMC analysis.
Focusing on DM particles with masses between MeV and GeV, 
we explore the parameter space of these three benchmark models, see Eq.~\eqref{eq:model1},~\eqref{eq:model2}, and~\eqref{eq:model3}, 
subject to the constraints summarized in Table~\ref{Tab:constraints}. 
We use \texttt{FeynRules}~\cite{Alloul:2013bka} to implement the Lagrangian and \texttt{CalcHEP}~\cite{Belyaev:2012qa} to calculate the cross sections.

For a Gaussian likelihood, we define the chi-square as 
\begin{equation}
\chi^2 = \frac{(\mu_{\text{theo}} - \mu_{\text{exp}})^2}{\sigma_{\rm tot.}^2}
\label{eq:likelihoods}
\end{equation}
where $\mu_{\text{theo}}$ represents the theoretical prediction and $\mu_{\rm exp}$ denotes the experimental central value. 
The total uncertainty $\sigma_{\rm tot.}$ is given by $\sigma_{\rm tot.} = \sqrt{\sigma_{\rm theo}^2 + \sigma_{\rm exp}^2}$. 
For a Half-Gaussian likelihood, where no signal is expected, we set $\mu_{\rm exp} = 0$. 
In the following, we outline relevant constraints in Table~\ref{Tab:constraints} and their associated likelihood functions.

\begin{itemize}
    \item \underline{Relic abundance}\\
We use the Boltzmann solver \texttt{MicrOMEGAs}~\cite{Belanger:2020gnr} to compute the predicted DM relic density $\Omega_{\rm DM}^{\rm theo} h^2$. 
We implement a Gaussian likelihood function adopting the central value and statistical error from the Planck 2018 measurement~\cite{Planck:2018vyg}. 
In addition, we include a systematic uncertainty of $\sigma_{\rm sys}=10\%$ in the relic density calculation. 
This accounts for theoretical uncertainties associated with the effective number of relativistic degrees of freedom, $g_\star(T)$ and $g_{*S}(T)$, entering the Boltzmann equation during freeze-out. 
In the sub-GeV regime, freeze-out can occur near the QCD phase transition, where non-perturbative effects lead to $\mathcal{O}(10\%)$ variations in $g_\star$ and $g_{*S}$ among different evaluations~\cite{Drees:2015exa}.
Such differences translate directly into comparable shifts in the predicted relic density. 
To remain conservative, we assume the maximal $\sim 10\%$ uncertainty reported in the literature~\cite{Fowlie:2013oua, Kowalska:2014hza,Saikawa:2020swg,Watanabe:2025pvc}. 

    \item \underline{DM direct detection}\\
For sub-GeV DM particles scattering with nuclei, the low energy deposition ($v_{\rm DM} \approx 10^{-3}$) can only be detected in lower-threshold experiments, 
such as DarkSide-50~\cite{DarkSide-50:2023fcw} and CRESST-III~\cite{CRESST:2019jnq,CRESST:2022lqw}, which are most sensitive in the DM mass range $60\ \text{MeV} < m_{\rm DM} < 1\ \text{GeV}$ for DM-nucleon interactions. 
Leptophilic interactions do not involve tree-level DM-quark elastic scattering, but DM can couple to nucleons via photon and lepton loops. 
We use the general lepton current ($\bar{l} \Gamma_l l$) to calculate the theoretical spin-independent DM-nucleon scattering cross section $\sigma_{p}^{\text{SI}}$.

For a scalar lepton current ($\Gamma_l = 1$), the scalar current does not couple to a vector current, 
thus DM-quark interactions arise only at the two-loop level for both $\mathcal{L}_{SS}$ and $\mathcal{L}_{FS}$. 
We compute $\sigma_{p}^{\text{SI}}$ using Eq.~(B1) and (B5) from Ref.~\cite{Abdughani:2021oit}.

For axial-vector lepton currents ($\Gamma_l = \gamma_\mu \gamma_5$), 
the corresponding diagrams vanish at all loop orders. 
Therefore, we ignore DM-nucleon scattering constraints for $\mathcal{L}_{FV}$.

In addition to DM-nucleon scattering, sub-GeV DM can scatter off electrons, producing detectable single-electron ionization signals. 
However, recent results from XENON1T~\cite{XENON:2019gfn}, DarkSide-50~\cite{DarkSide:2022knj}, DAMIC-M~\cite{DAMIC-M:2023gxo}, and SENSEI~\cite{SENSEI:2020dpa} place stringent limits on such interactions. 
While electrophilic DM couples to electrons at tree level, muonphilic DM lacks this interaction and relies on suppressed loop-level contributions. 
As no DM-electron scattering signal has been observed, these experiments impose strong constraints on the parameter space of electrophilic DM.

We compute the DM-free electron scattering cross section $\bar\sigma_e$ using the \texttt{CalcHEP}~\cite{Belyaev:2012qa} package. 
Assuming a direct coupling between DM and electrons, we adopt the model-independent parameterization from Ref.~\cite{Essig:2011nj}, 
with the reference cross section defined as
\begin{equation}
    \begin{aligned}
        & \overline{\sigma}_e = \frac{\mu_{{\rm DM}e}^2}{16\pi m_{\rm DM}^2 m_e^2} 
        \left. \overline{\left|\mathcal{M}_{{\rm DM}e}(\alpha m_e)\right|^2} \right., \\
        & {\rm where} \quad
        \left.\overline{\left|\mathcal{M}_{{\rm DM}e}(q)\right|^2} \right. = 
        \left. \overline{\left|\mathcal{M}_{{\rm DM}e}(\alpha m_e)\right|^2} \right. 
        \times \left|F_{\rm DM}(q)\right|^2.
    \end{aligned}
\end{equation}
The squared matrix element averaged over initial and summed over final spins with 3-momentum transfer $q$ is denoted as 
$\overline{\left|\mathcal{M}_{{\rm DM}e}(q)\right|^2}$. 
For models with a light mediator ($m_{\rm{MED}} \ll \rm{keV}$), the form factor is $F_{\rm{DM}}(q) = (\alpha m_e / q)^2$~\cite{Essig:2011nj, Essig:2019xkx}. 
In our case, with $m_{\rm{MED}} \gg \alpha m_e$, we take $F_{\rm{DM}}(q) = 1$.

To better illustrate the impact of DM-electron scattering constraints, 
we do not include these limits in the likelihood function, since they primarily affect electrophilic DM.
Instead, we display them directly in the $(m_{\rm DM}, \bar{\sigma}_e)$ plane, 
where their effect on the excluded parameter space is more clearly visualized.
Detailed results and discussions are presented in Sec.~\ref{sec:result}.

    \item \underline{CMB}\\
DM can annihilate either directly into two leptons ($s$-channel) or into four leptons through mediator decays ($t$-channel). CMB anisotropies are sensitive only to the total electromagnetic energy deposited in the thermal plasma, which can be parameterized by the efficiency factor $f_{\rm eff}$ multiplying $\langle \sigma v \rangle/m_{\rm DM}$. While in principle $f_{\rm eff}$ differs somewhat between two-body and four-body leptonic final states, the variations are $\mathcal{O}(1)$~\cite{Slatyer:2015jla} and do not qualitatively affect the likelihood function in the sub-GeV mass range relevant to this work. 
Therefore, we simply apply the updated Planck 2018 limits~\cite{Planck:2018vyg} to both channels.

The DM annihilation cross section during the CMB era is denoted as $\sv_{\text{CMB}}$, and its velocity-average $\langle \sigma v \rangle_{\rm{CMB}}$ is 
    \begin{eqnarray}
     \left \langle \sigma v  \right \rangle_{\rm{CMB}}&=&\int \sigma v f(v)\mathrm{d}v,  \nonumber\\    
   {\rm where}~f(v)&=& \frac{v^{2} }{2\sqrt{\pi }v_{\rm CMB}^{3}} \exp\left[{-\frac{v^{2} }{4v_{\rm CMB}^{2} } }\right].
    \end{eqnarray}
The annihilation cross section $\sigma v$ is computed using the \texttt{CalcHEP}~\cite{Belyaev:2012qa}.
The DM velocity at recombination epoch $v_{\rm CMB}$ is derived as 
    \begin{equation}
    v_{\rm CMB}^2=3 x\left(\frac{T_{\gamma}^{\rm CMB}}{m_{\rm{DM}}}\right)^2,
    \end{equation}
where the dimensionless temperature parameter $x\equiv m_{\rm DM}/T$ is determined as the maximum of the freeze-out temperature parameter $x_F$ and the kinematic decoupling temperature parameter $x_{\rm kd}$.

For CMB likelihood function, we use a Half-Gaussian distribution with Chi-squared defined as 
    \begin{equation}
    \chi_{\text{CMB}}^2=\left[\frac{\sv_{\text{CMB}}}{\sv^{\text{Planck},95\%}_{\text{CMB}}/1.64}\right]^2 ,
    \label{eq:chi2_CMB}
    \end{equation}
    where $\sv^{\text{Planck},95\%}_{\text{CMB}}$ is the upper limit on the cross section for a given DM mass at the $95\%$ confidence level, as reported by Planck 2018 data~\cite{Planck:2018vyg}.

    \item \underline{DM self interaction}\\
Since $g_f$ can be suppressed by DM direct detection or light mediator searches, 
$g_D$ shall be sizable to ensure the correct relic abundance. 
As a result, with a large $g_D$, the DM self-interaction cross section $\sigma_{\text{DM,DM} \to \text{DM,DM}}$ may become significant, 
potentially exceeding the upper limit set by the Bullet Cluster, $\sigma_{\text{DM,DM} \to \text{DM,DM}} / m_{\text{DM}} < 1.0 \, \text{cm}^2/\text{g}$~\cite{Randall:2008ppe}, 
where the DM incident velocity is $4000\,\text{km/s}$.

    \item \underline{$N_{\rm eff}$}\\
A light mediator may contribute to the relativistic degrees of freedom $N_{\rm eff}$.
We incorporate this effect by implementing a half Gaussian likelihood function with a $95\%$ confidence level upper limit $\Delta N_{\rm eff}< 0.17$~\cite{Planck:2018vyg}.
This constraint effectively rules out the parameter space where both $m_{\rm MED}$ and $m_{\rm DM}$ are below a few MeV.

    \item \underline{BBN}\\
Mediator decay into SM particles may disrupt BBN thermal history. 
In this work, we assume the mediator decays only into leptons. 
To secure a successful BBN thermal history, we impose the conservative condition $\tau_{\text{MED}} \le 10^5 \, \text{s}$~\cite{Krnjaic:2015mbs}.

    \item \underline{Mediator production}\\
For mediator particles in the MeV to GeV mass range, the most sensitive beam-dump experiments for detecting leptonphilic mediator production are NA64~\cite{2502.04053, 2409.10128}, Belle II~\cite{2403.02841}, and BABAR~\cite{1606.03501}. 
However, both theoretical models and experimental constraints are categorized based on electronphilic and muonphilic mediators. 
Since each experiment aims at different mediator types and employs distinct analysis methods, 
we apply conditional cuts to ensure consistency with their respective predictions and measurements.

For $m_{\rm MED}$ between $1\mev$ to $100\mev$, NA64$e$~\cite{2502.04053} searches for the bremsstrahlung process $e^\pm N \to {\rm MED}~e^\pm N$, 
where the mediator decays invisibly into DM pairs. 
Similarly, NA64$\mu$~\cite{2409.10128} probes mediator bremsstrahlung using a $\mu$ beam. 
These experiments are sensitive to our parameter space when the mediator decays invisibly, i.e., $2 m_{\rm DM} < m_{\rm MED}$. 
While they assume a 100\% branching ratio for invisible decays, this can vary in our parameter space.

For $m_{\rm MED} > 2 m_\mu$, the mediator can decay into muons, and Belle II~\cite{2403.02841} and BABAR~\cite{1606.03501} impose the strongest constraints. 
The mediator is produced via $e^+e^- \to {\rm MED}~\mu^+\mu^-$, and successively decays as ${\rm MED} \to 2\mu$. 
This process allows us to probe $g_f$ as a function of the mediator mass. 
For $m_{\rm MED} < 1\gev$, both BABAR and Belle II require $g_f \lesssim 10^{-3}$ at 95\% confidence level.

\end{itemize}

\section{Allowed parameter space}
\label{sec:result}

\begin{figure}[ht!]
	\centering
    \includegraphics[width=16.8cm,height=5.8cm]{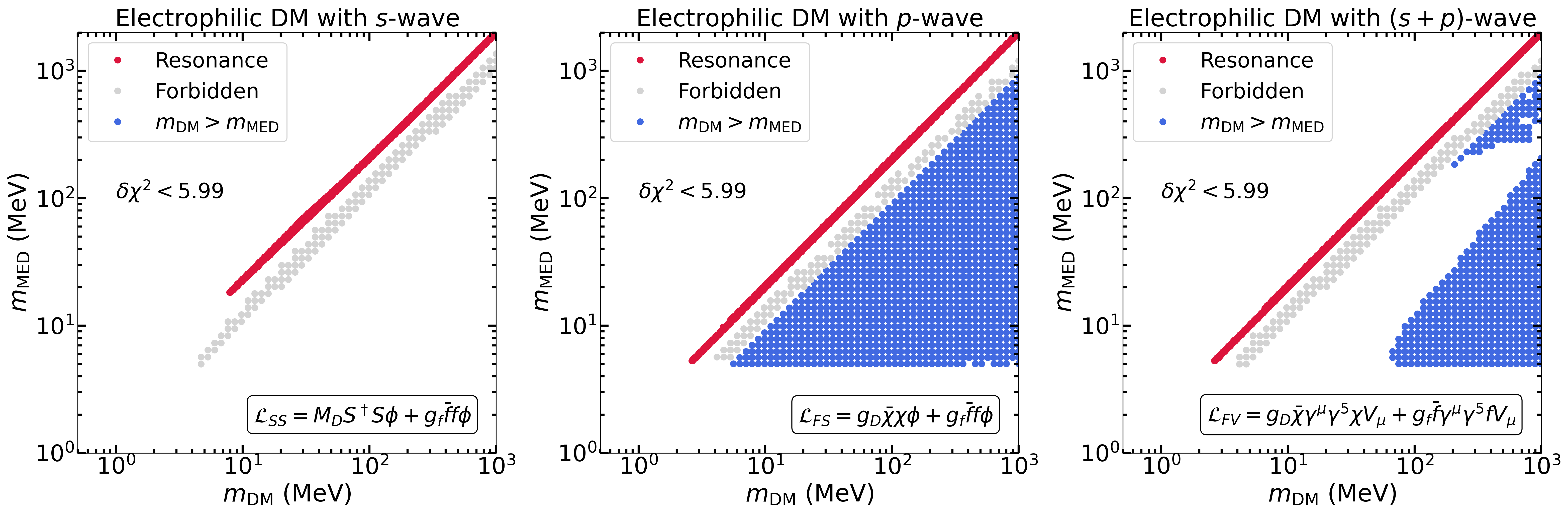}
	\includegraphics[width=16.8cm,height=5.8cm]{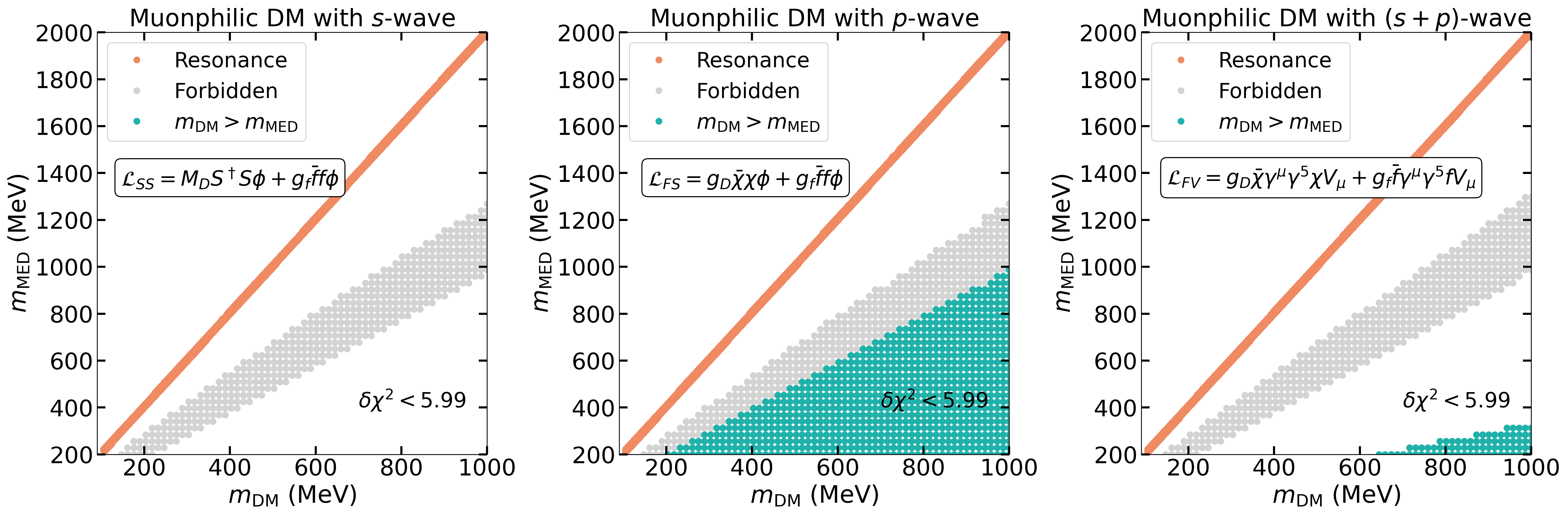}
\caption{The $2\sigma$ allowed regions ($\delta\chi^2 < 5.99$ for two degrees of freedom) 
in the $(m_\text{DM},m_\text{MED})$ plane for the three benchmark interactions. 
The upper panels correspond to the electrophilic DM scenario, 
while the lower panels correspond to the muonophilic DM scenario. 
Colored dots represent surviving parameter points from the MCMC scan. 
Red/orange indicate resonance regions, gray corresponds to forbidden annihilation, 
and blue/cyan mark the parameter space where $m_{\rm DM} > m_{\rm MED}$.
}
	\label{Fig:mx_mphi}
\end{figure}

For each model, there are four free parameters and their scan ranges are 
\begin{eqnarray}
	\{1\,\mev\,~{\rm or}~ m_\mu\}\leq &m_{\rm{DM}}& \leq\, 1\,{\rm GeV}, \nonumber \\
	\{1~\mev\,~{\rm or}~ 2 m_\mu\} \leq & m_{\rm{MED}} & \leq\, 2\,{\rm GeV}, \nonumber \\    
    10^{-6} \leq & \{g_D~{\rm or}~M_D/\gev\} & \leq\, 1,\nonumber \\
        10^{-6} \leq &g_f& \leq 1,
    \label{eq:parameter range}
\end{eqnarray}
where the lower limits of $m_{\rm{DM}}$ and $m_{\rm{MED}}$ vary to keep the $\mu^+\mu^-$ annihilation channel open in muonphilic models.
Note that the coupling constant $M_D$ has dimension one for $\mathcal{L}_{SS}$, while $g_D$ is dimensionless for $\mathcal{L}_{FS}$ and $\mathcal{L}_{FV}$.

We use \texttt{emcee}~\cite{Foreman-Mackey:2012any}, a MCMC sampler, to explore the parameter space with the likelihood functions from Sec.~\ref{sec:constraints}.  
We run 20 Markov chains for each model to ensure the coverage of the four-dimensional parameter space defined in Eq.~(\ref{eq:parameter range}).  
To present our results, we apply the Profile Likelihood method to eliminate nuisance parameters and project the results onto a two-dimensional plane.  
Assuming a nearly Gaussian likelihood, 
we adopt the $2\sigma$ (95\% C.L.) allowed region as $\delta\chi^2 < 5.99$ for two degrees of freedom.

Fig.~\ref{Fig:mx_mphi} shows the $2\sigma$ allowed region in the $(m_{\rm DM}, m_{\rm MED})$ plane, 
with electrophilic and muonophilic DM depicted in the upper and lower panels, respectively. 
The colored dots represent parameter points satisfying total Chi-squared $\delta\chi^2 < 5.99$. 
Colors distinguish the surviving points based on their dominant annihilation mechanisms: 
red/orange for resonance, gray for forbidden regions, and blue/cyan for $m_{\rm DM} > m_{\rm MED}$. 
Specifically, red and blue denote the electrophilic DM (upper panels), while orange and cyan correspond to muonophilic DM (lower panels).
The red and orange regions indicate resonance annihilation ($m_{\rm MED} \approx 2 m_{\rm DM}$), where DM satisfies thermal relic density and CMB limits, independent of $s$-, $p$-, or $(s+p)$-wave processes.
However, CMB data exclude the resonance $s$-wave annihilation parameter space for $m_{\rm DM} < 10\,\mathrm{MeV}$, as seen in the upper left panel.
Additionally, in electrophilic scenarios, the lower bound on $m_{\rm{MED}}$ and $m_{\rm{DM}}$ are set by constraints from $\Delta N_{\rm eff}$ measurements.

The gray region corresponds to the kinematic threshold regime where $m_{\rm MED} \approx m_{\rm DM}$. In the early universe, when DM was semi-relativistic, annihilation into mediator pairs was possible and could efficiently deplete the abundance to yield the observed relic density. In contrast, in the present-day Galactic environment where DM is non-relativistic ($v\sim 10^{-3}$), the available kinetic energy is far too small to overcome the mass threshold. As a result, annihilation into mediator pairs is forbidden today, and this parameter region does not lead to observable indirect detection signals, even in dense DM systems such as dwarf spheroidal galaxies.

For $s$-wave annihilation, the cross section is constant, resulting in stringent CMB constraints and leaving only the resonance and forbidden regions viable. 
In contrast, $p$-wave annihilation escapes CMB constraints due to velocity suppression, allowing a broader parameter space. 
The parameter space with $m_{\rm{DM}} > m_{\rm{MED}}$ becomes particularly interesting for the $(s+p)$-wave mechanism. 
We can parameterize the velocity dependence of the annihilation cross section as $\sv \simeq a + b v^2$, assuming lower velocities. 
Ref.~\cite{Abdughani:2021oit} shows that in a vector mediator scenario characterized by the Lagrangian $\mathcal{L}_{FV}$, 
the cross section $\sv_{2f}$ for the process ${\rm DM}\,{\rm DM} \to f\bar{f}$ yields a ratio $a/b \approx \mathcal{O}(1)$. 
Given the extremely low DM velocities during the CMB epoch ($v \sim 10^{-6}$ to $10^{-8}$), the cross section $\sv_{2f}$ is dominated by the $s$-wave component, which excludes the parameter space $m_{\rm{DM}} \gtrsim m_{\rm{MED}}$ based on CMB data. 
In contrast, for the process ${\rm DM}\,{\rm DM} \to {\rm MED}\,{\rm MED} \to 4f$, the ratio $a/b \ll 1$ indicates that the cross section $\sv_{4f}$ behaves like a $p$-wave.

\begin{figure}[htb!]
	\centering
    \includegraphics[width=16.8cm,height=5.8cm]{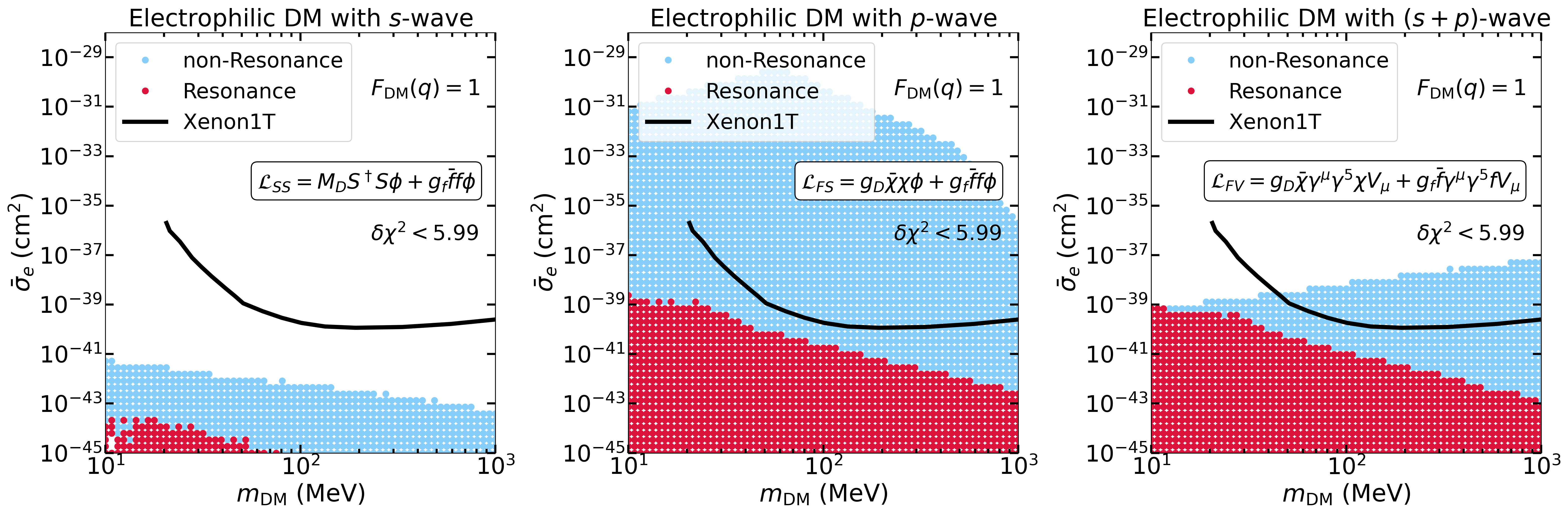}
\caption{The $2\sigma$ allowed region for the DM-electron scattering cross section $\bar\sigma_e$ as a function of $m_\text{DM}$. 
The region labeled as non-resonance region includes the forbidden and $m_{\rm DM}>m_{\rm MED}$ scenarios, 
as defined in Fig.~\ref{Fig:mx_mphi}. 
The form factor is taken as heavy mediator limit, $F_{\rm DM}(q)=1$.
The region above the solid lines is excluded by the null results from the XENON1T~\cite{XENON:2019gfn}, DarkSide-50~\cite{DarkSide:2022knj}, and DAMIC-M~\cite{DAMIC-M:2023gxo} DM search experiments. 
}
	\label{Fig:DD}
\end{figure}

Considering direct detection for DM-electron interactions, the electrophilic and muonphilic scenarios yield very different predictions. 
In the electrophilic case, DM couples to electrons through a tree-level diagram, which significantly enhances the DM-free electron scattering cross section $\bar\sigma_e$, making it accessible for current direct detection experiments. 
In contrast, in the muonphilic case, the absence of a tree-level contribution to $\bar\sigma_e$ leads to substantial suppression from loop-level processes.

In Fig.~\ref{Fig:DD}, we show the $2\sigma$ allowed region projected onto the ($m_\text{DM}$, $\bar\sigma_e$) plane for electrophilic interactions.
The light blue regions indicate the non-resonance regime, which includes the forbidden and $m_{\rm DM} > m_{\rm MED}$ regions (as shown in Fig.~\ref{Fig:mx_mphi}). Our scan for the mediator mass is above $1\,\rm{MeV}$, specifically $m_{\rm{MED}} \gg \alpha m_e$, allowing us to use the heavy mediator form-factor $F_{\rm{DM}}(q)=1$. 
It is worth noting that most of the non-resonance region in both the $p$-wave (middle panel) and $(s+p)$-wave (right panel) scenarios have already been excluded by the null results from XENON1T~\cite{XENON:2019gfn}, DarkSide-50~\cite{DarkSide:2022knj}, and DAMIC-M~\cite{DAMIC-M:2023gxo}, while the resonance region exhibits a suppressed $\bar\sigma_e$, making it difficult to be probed with these experiments.
This suppression arises from the Planck relic density constraint, which imposes an upper limit on the coupling constants $g_D$ and $g_f$.
In the following, we will focus exclusively on the resonance region, which can bypass the constraints imposed by relic density, CMB, and DM direct detection.

\begin{figure}[htb]
	\centering
    \includegraphics[width=16.8cm,height=5.8cm]{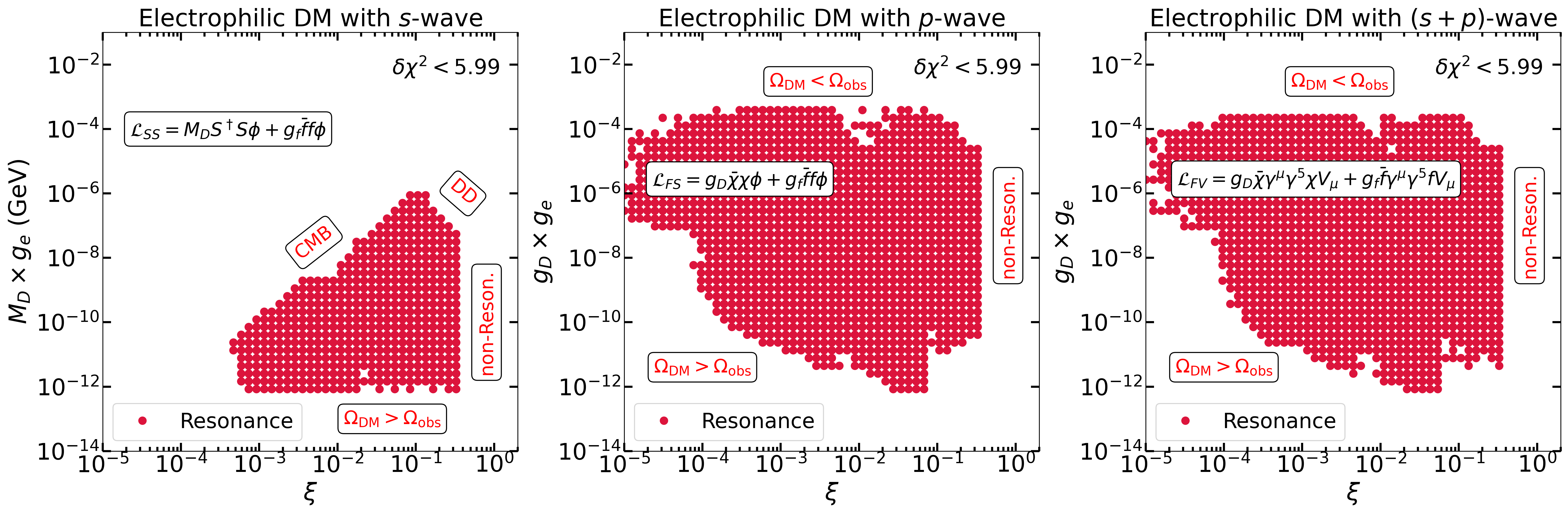}
	\includegraphics[width=16.8cm,height=5.8cm]{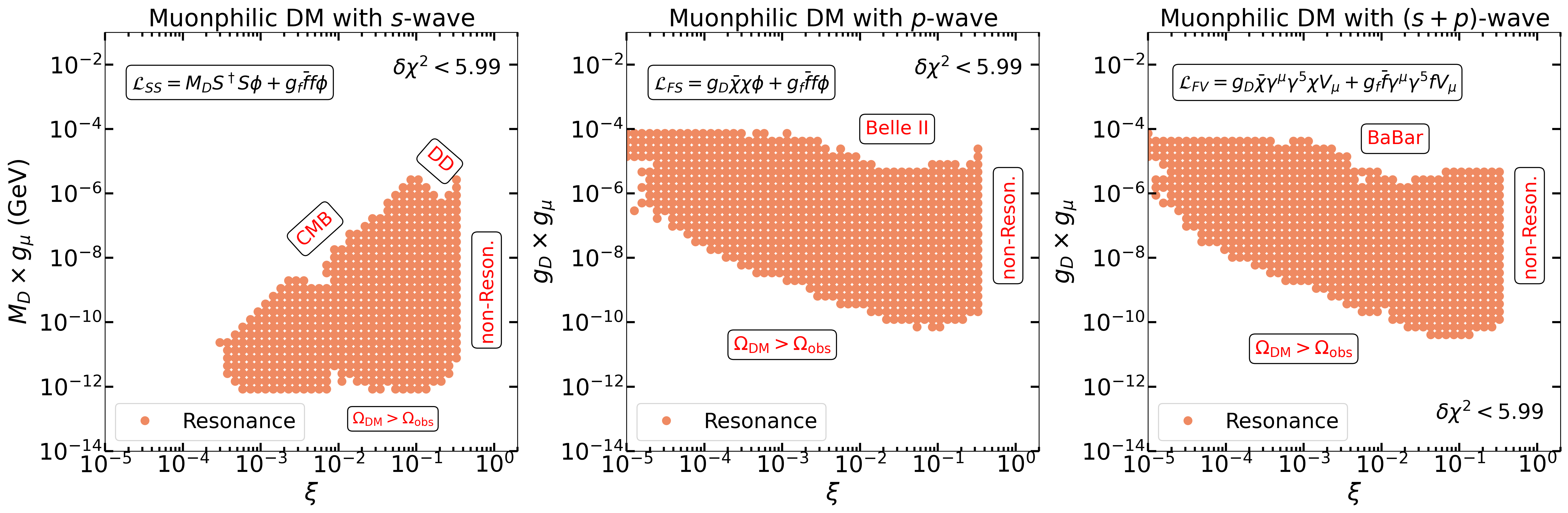}
\caption{The $2\sigma$ allowed regions for the resonance regions.  
The coupling product $M_D \times g_{e\,(\mu)}$ (left column) and $g_D \times g_{e\,(\mu)}$ (middle and right columns) are the function of the resonance parameter $\xi$ in the resonance regime. The resonance parameter is defined as $\xi=\sqrt{(m_{\rm{MED}}^2-4m_{\rm{DM}}^2)/4m_{\rm{DM}}^2}$.  
}
	\label{Fig:couplings}
\end{figure}

To illustrate the impact of all constraints on the resonance parameter space, we present in Fig.~\ref{Fig:couplings} the coupling product $M_D \times g_{e\,(\mu)}$ (left column) and $g_D \times g_{e\,(\mu)}$ (middle and right columns) as a function of the resonance parameter $\xi$, 
defined as $\xi = \sqrt{(m_{\rm{MED}}^2 - 4m_{\rm{DM}}^2)/4m_{\rm{DM}}^2}$, within the $95\%$ confidence region. 
Additionally, we label the exclusion regions with appropriate experimental constraint tags, marked in red. 
We summarize our findings as follows:
\begin{itemize}
    \item For the $s$-wave annihilation, $M_D \times g_{e(\mu)}$ exhibits a approximate linear decrease with $\xi$, 
    a behavior resulting from stringent CMB and relic abundance constraints. 
    In the parameter region $\xi > 1 \times 10^{-3}$, direct detection experiments impose strong constraints on $M_D \times g_{\mu}$.
    
    \item In the muonphilic scenario, the upper limits of $g_D \times g_\mu$ for the $p$-wave and $(s+p)$-wave mechanisms are set by Belle II and BaBar, respectively.
    As shown in Ref.~\cite{2403.02841}, BaBar provides slightly stronger constraints than Belle II in the case of a vector mediator.
    We would like to note that NA64e~\cite{2502.04053} and NA64$\mu$~\cite{2409.10128} do not improve the upper limits for either the electrophilic or muonphilic resonance region, 
    as these experiments are only sensitive to our parameter space when the mediator decays invisibly, i.e., in the non-resonance region.
    
    \item The observed DM relic density provides a lower limit on coupling products, 
    as too small coupling constants can result in DM overabundance. 
    
    \item The upper bound on $\xi$ is determined by the resonance condition. 
    Ref.~\cite{Wang:2025tdx} suggests that $\xi^2$ can be as small as $\mathcal{O}(10^{-16})$, but achieving such a small value would require extreme fine-tuning, which is beyond the scope of this work.
\end{itemize}

\section{Probing resonance region by VLAST}
\label{sec:VLAST_result}

\begin{figure}[htb!]
	\centering
    \includegraphics[width=16.8cm,height=5.8cm]{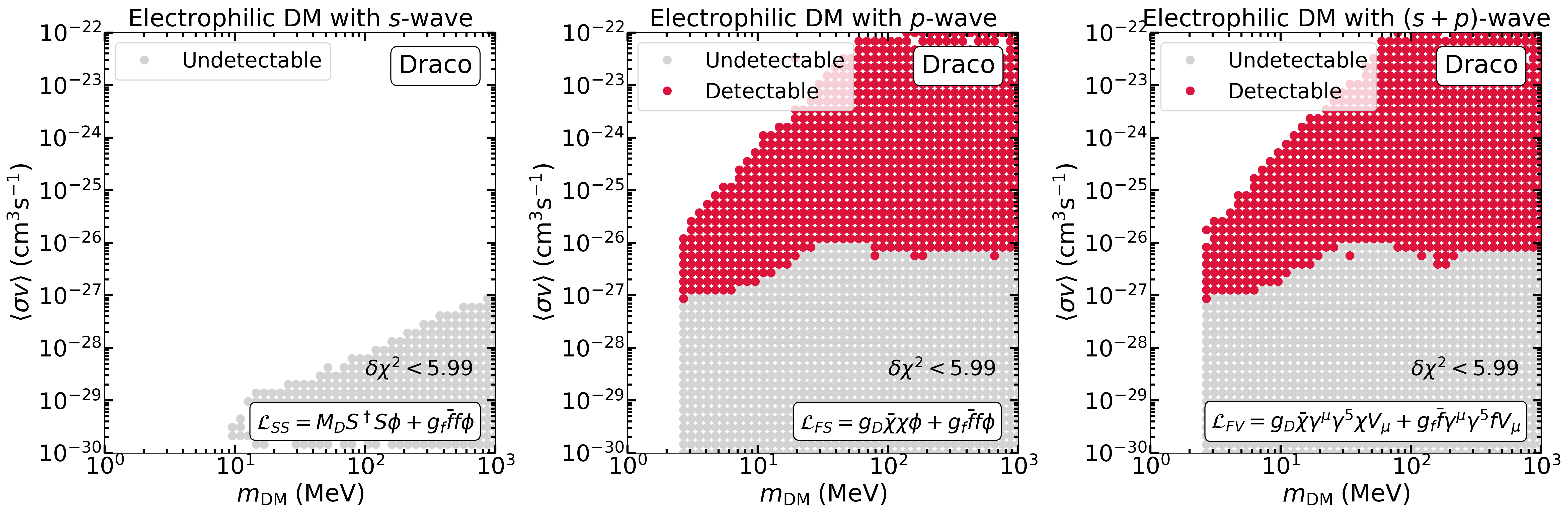}	\includegraphics[width=16.8cm,height=5.8cm]{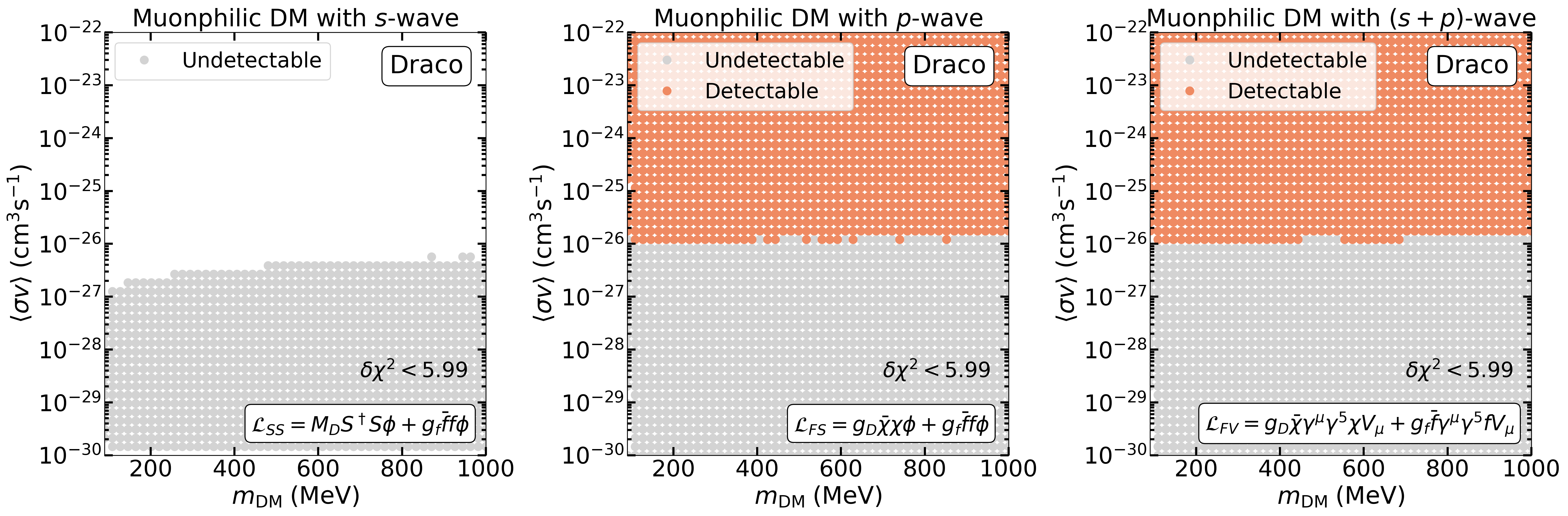}
\caption{Expected sensitivity of VLAST to DM annihilation via mediator resonance in Draco dwarf galaxy over a 5-year observation period. 
The red and orange regions denote the parameter space accessible to VLAST for the electrophilic (upper panels) and muonphilic (lower panels) DM, respectively, while the gray region represents the parameter space that remains undetectable.}
	\label{Fig:sigmav}
\end{figure}

In Fig.~\ref{Fig:sigmav}, we show the projected sensitivity of VLAST to DM annihilation via mediator resonance within Draco dwarf galaxy over a 5-year observation period, presenting three different interactions across three columns for electrophilic (upper panels) and muonphilic (lower panels) DM. 
The red and orange regions indicate the parameter space where VLAST is sensitive to DM annihilation signals, while the gray region represents the parameter space where the annihilation cross section is too low for the VLAST future detection.

For annihilation via mediator resonance in the $p$-wave and $(s+p)$-wave cases, VLAST is expected to provide strong sensitivity to DM annihilation signals.
The white regions in the electrophilic $p$-wave (upper middle panel) and $(s+p)$-wave (upper right panel) correspond to parameter points that are excluded by existing constraints rather than being undetectable.
At very low masses ($m_{\rm DM}\lesssim 3\,\rm{MeV}$), the lower boundary is set by the $\Delta N_{\rm eff}$ constraint.
For DM masses between roughly $3\,\rm{MeV}$ and $55\,\rm{MeV}$, parameter points with large annihilation cross sections are excluded by CMB constraints, as the Planck limits on energy injection become increasingly stringent toward lighter DM masses.

VLAST exhibits limited sensitivity to the $s$-wave scenario, even with the resonance mechanism. 
This limitation arises from the combination of the DM relic abundance and CMB constraints: achieving the correct relic density requires a sufficiently large annihilation cross section, while evading stringent CMB constraints needs a small cross section during the CMB epoch. 
The only way to meet both conditions in the $s$-wave case is for the resonance peak to be near the freeze-out epoch, which generates a suppression of the annihilation cross section at later times when the DM velocity is sufficiently lower.
Consequently, the white regions in the $s$-wave panels (upper and lower left) correspond to parameter points excluded by this tension between relic density and CMB constraints.
In Draco dwarf galaxy, where the typical DM relative velocity is approximately $\sim 10^{-4}$ in the unit of light speed, this suppression persists. 
Therefore, DM annihilation with $s$-wave resonance remains undetectable.

\section{Summary and conclusion}
\label{sec:summary}

VLAST is expected to advance our knowledge of gamma-ray astrophysics, particularly in addressing the long-standing ``MeV Gap'' in indirect DM detection. With its superior sensitivity in the MeV to GeV energy range and enhanced energy resolution, 
VLAST offers a unique opportunity to probe the thermal relic annihilation signals of sub-GeV DM, especially in DM-rich system such as dwarf spheroidal galaxies.

In this work, we have examined the VLAST sensitivity to three representative interactions of leptophilic DM: 
(i) scalar DM with a scalar mediator ($s$-wave annihilation), 
(ii) Dirac DM with a scalar mediator ($p$-wave annihilation), and 
(iii) Dirac DM with a vector mediator (mixed $s$- and $p$-wave). 
We have individually discussed these interactions with electrophilic and muonphilic annihilation channels and 
kinematic behaviors expected in the sub-GeV mass regime.

Our comprehensive analysis incorporates several experimental constraints, including CMB anisotropies from Planck, relic abundance from cosmological observations, direct detection bounds (e.g., XENON1T and DarkSide-50), self-interaction limits from astrophysical probes (e.g., the Bullet Cluster), BBN requirements, and bounds from beam-dump experiments such as NA64, Belle II, and BABAR.

A central focus of our study is the \textit{resonance region} (where $m_{\text{MED}} \approx 2m_{\text{DM}}$), 
which allows for enhanced annihilation cross sections during DM freeze-out without violating CMB constraints in the present epoch. 
This resonance condition enables viable $s$-wave, $p$-wave, and mixed $(s+p)$-wave annihilation, 
although the parameter space remains heavily constrained for the $s$-wave case due to its velocity-independent behavior.

VLAST is shown to be particularly sensitive to the \textit{resonance $p$-wave and $(s+p)$-wave} regions. 
These scenarios benefit from velocity suppression in the early universe (evading CMB limits) while still permitting detectable annihilation signals in galaxies with relatively higher DM velocities today (e.g., $v \sim 10^{-4}$). 
In contrast, $s$-wave scenarios, even in resonance, remain largely undetectable by VLAST due to the suppression required to satisfy relic and CMB constraints simultaneously.

We further distinguish between \textit{electrophilic} and \textit{muonphilic} interactions. 
In the electrophilic case, direct detection experiments impose strong constraints on the non-resonance parameter space due to tree-level DM-electron scattering. However, VLAST can effectively probe the resonance region that escapes these bounds. 
In contrast, the muonophilic case avoids stringent direct detection limits by suppressing loop-level interactions with electrons, 
making the muon channel an appealing annihilation final state for future indirect searches.

Overall, our study shows that VLAST has the potential to significantly advance the search for light leptophilic DM. 
Its superior sensitivity between MeV to GeV energy range enables it to explore viable regions of the parameter space that remain inaccessible to both direct detection and current gamma-ray observatories. 
Particularly for $p$-wave and $(s+p)$-wave annihilation scenarios in the resonance region, VLAST emerges as a critical instrument for probing the thermal relic nature of sub-GeV DM.

\section*{Acknowledgments}
We were supported by the National Key Research and Development Program of China (No. 2022YFF0503304), and 
the Project for Young Scientists in Basic Research of the Chinese Academy of Sciences (No. YSBR-092). 
TPT is also supported by the Jiangsu Province Post Doctoral Foundation (No. 2024ZB713).
\newpage

\appendix

\bibliography{ref}


\end{document}